# A Starshot Communication Downlink

Kevin L.G. Parkin[a]

[a] *Systems Director, Breakthrough Starshot. Parkin Research LLC, 2261 Market St. #221*, San Francisco, California 94114, USA, kevin@parkinresearch.com

**Abstract**

Breakthrough Starshot is an initiative to propel a sailcraft to Alpha Centauri within the next generation. As the sailcraft transits Alpha Centauri at 0.2 c, it looks for signs of life by imaging planets and gathering other scientific data. After the transit, the 4.1-meter diameter sailcraft downlinks its data to an Earth-based receiver. The present work estimates the raw data rate of a 1.02 μm, 100 Watt laser that is received at 1.25 μm by a 30-meter telescope. The telescope receives 288 signal photons per second (-133 dBm) from the sailcraft after accounting for optical gains (+296 dBi), conventional losses (-476 dB), relativistic effects (-3.5 dB), and link margin (-3.0 dB). For this photon-starved Poisson channel with 0.1 nm equivalent noise bandwidth, 90% detector quantum efficiency, 1024-ary PPM modulation, and $10^{-3}$ raw bit error rate, the raw data rate is 260 bit/s (hard-decision) to 1.5 kbit/s (ideal) raw data rate, which is 8-50 Gbit/year. This rate is slowed by noise, especially starlight from Alpha Centauri A scattered into the detector by the atmosphere and receiver optics as sailcraft nears the star. Because this is a flyby mission (the sailcraft does not stop in the Centauri system), the proper motion of Alpha Centauri relative to Earth carries it away from the sailcraft after transit, and the noise subsides over days to weeks. The downlink can resume as soon as a day after transit, starting at 7-22 bit/s and reaching nearly full speed after 4 months. By using a coronagraph on the receiving telescope, full-rate downlink speed could be reached much sooner.

**Nomenclature**

- $\eta$    Detector quantum efficiency
- $G_R$    Receiver gain
- $G_T$    Transmitter gain
- $\gamma$    Photons (used as a unit)
- $I$    Irradiance
- $L_\beta$    Relativistic loss
- $L_R$    Conventional losses, including path loss, atmospheric transmission losses and link margin, but not relativistic loss
- $M$    Number of time slots per PPM frame, equal to peak-to-average power ratio
- $n_s$    Average number of signal photons detected per pulse, equal to $S\eta M T_s$
- $n_n$    Average number of noise photons detected per slot, equal to $N\eta T_s$
- $N$    Received noise power
- $P_T$    Transmitter input power
- $S$    Received signal power
- $T_s$    Time duration of a slot in a PPM frame
- $\theta$    Angular separation between star and detector

**Acronyms/Abbreviations**

- Alpha Centauri (αCen)
- Difference in right ascension (ΔRA)
- Difference in declination (ΔDec)
- Diffuse Infrared Background Experiment (DIRBE)
- Pulse position modulation (PPM)

## 1. Introduction

Breakthrough Starshot is an initiative to propel a sailcraft to αCen within the next generation. As the sailcraft passes αCen, it looks for signs of life by imaging planets and gathering other scientific data. After the flyby, the sailcraft's communication downlink returns the data to an Earth-based receiver.

*1.1 Approach*

Starshot proceeds from the determination that relativistic laser-driven sails do not violate known physics. Starshot does not proceed from the determination that relativistic laser-driven sails are within reach of the current art of engineering. Some mission requirements are expected to need basic research in engineering and applied physics. The Starshot timeframe is correspondingly long.

*1.2 Downlink subsystem status*

Currently, Starshot is collecting downlink solutions that meet mission requirements and do not violate known physics. The present work is one such candidate.

*1.3 Motivation for the present work*

To build a communication downlink model and link budget, as is typical in conceptual space mission design. To explore the potential performance of a near-infrared laser downlink. In future, to use the model in conjunction with the Starshot system model [1] to derive technology performance thresholds and subsystem requirements for optical downlink mission candidates.

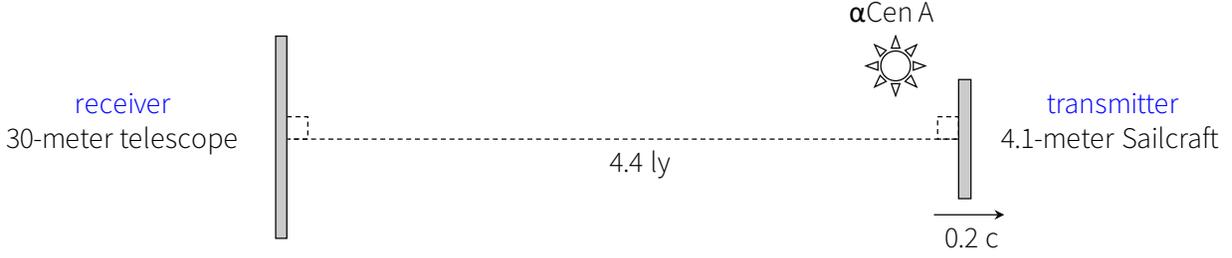

Fig. 1. Arrangement of the transmitter relative to the receiver for data downlink following transit of αCentauri A

*1.3 Prior work*

Prior work [2-5] has already estimated the performance of optical downlinks from αCen. This work draws on the prior work but starts from different assumptions that are consistent with the Starshot mission as it is currently understood [1].

**2. Assumptions**

The receiver is a 30-meter diffraction-limited telescope with 0.1 nm equivalent noise bandwidth at 1.25 μm. To simplify the analysis, the receiver is fixed in the position and direction relative to the transmitter shown in Fig. 1.

The transmitter is a 4.1-meter diameter sailcraft. Having recently passed αCen A, the sailcraft now exactly faces the receiver at a distance of 4.4 ly and recedes from it at 0.2 c.

Crucially, the entire sailcraft acts as a primary optic that forms near-ideal wavefronts across a 4.1-meter diameter plane facing the receiver. The wavefronts constitute a diffraction-limited infrared downlink beam with 70% aperture efficiency. The sailcraft transmits 100 Watts, deriving its power from a 700-Watt hydrogen beam that is normally incident on the sailcraft when it faces Earth [1]. This hydrogen beam is simply a manifestation of the interstellar medium, incident on the sailcraft at 0.2 c. Such aperture and power assumptions are beyond the current art of engineering for a sailcraft as currently envisaged. However, it is necessary to choose these assumptions in order to discover what the downlink performance of a fully-realized sailcraft can be.

**3. Models**

The downlink model is formed by joining a signal model and noise models with a channel model, as depicted by the Euler diagram in Fig. 2.

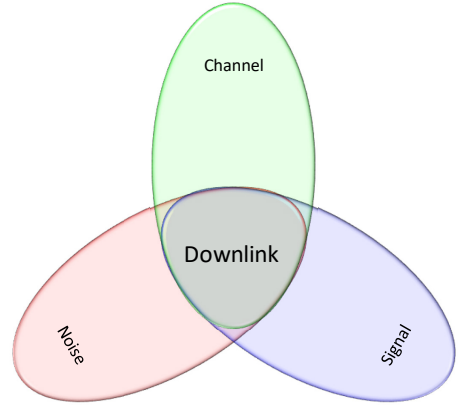

Fig. 2. The downlink model

*3.1 Signal*

The received signal is given by a relativistic version of the Friis transmission equation,

$$S[\text{dBm}] = P_T[\text{dBm}] + G_T[\text{dBi}] + G_R[\text{dBi}] + L_R[\text{dB}] + L_\beta[\text{dB}]. \quad (1)$$

In this relativistic case, there is an extra term $L_\beta$ that accounts for relativistic dimming. For the arrangement shown in Fig. 1, it is equal to the fourth power of the Doppler factor [6]. In compiling the link budget, shown in Table 1, quantities with subscript T are evaluated in the transmitter rest frame, and quantities with subscript R are evaluated in the receiver rest frame.

Table 1. Link budget

| | | |
|---|---|---|
| $P_T$ | +50 dBm | 100 W at 1.02 μm |
| $G_T$ | +140 dBi | 4.1 m diameter circular primary, 70% aperture efficiency |
| $G_R$ | +156 dBi | 30 m diameter circular primary, 70% aperture efficiency |
| $L_R$ | -476 dB | free-space path loss over 4.367 ly, 80% atmospheric transmittance, 3 dB link margin |
| $L_\beta$ | -3.5 dB | transmitter recedes from receiver at 0.2 c; Doppler effect, headlight effect |
| $S$ | -133 dBm | 288 photons/second at 1.25 μm |

The key finding of the link budget is that 288 photons per second reach the detector, meaning that the communication channel operates in a photon-starved regime.

### 3.2 Noise

The noise model sums the noise components traditionally contributed by the Earth's sky, reflected and re-radiated starlight from αCen A's dust disc, and direct light that is scattered into the detector by the telescope itself.

The sky and dust noise sources occupy the whole effective solid angle of the telescope, whereas the direct light

#### 3.2.1 Sky

The ESO SkyCalc Sky Model Calculator [7-9] is used to calculate sky noise radiance components at the telescope site.

Subject to the 'middle of the road' inputs in Table 2, the model generates the outputs in Table 3 and the radiance components shown in Fig. 3. From Paranal, αCen is only 32 degrees above the horizon, yet the atmosphere still transmits more than 80% of the photons at 1.25 μm.

The outputs of course vary depending on time, date, and location. A more detailed analysis would follow geographically-diverse receivers through all times and weather conditions to ensure a low probability of missed data.

Table 2. SkyCalc inputs

| | |
|---|---|
| Wavelength | 1.25 μm |
| Location | Cerro Paranal (2640 m altitude) |
| Target | αCen |
| Date | 19/01/2039 |
| Time | 06:55:34 UT (middle-of-road conditions) |
| Precipitable water vapor column | 2.5 mm (median for this site) |

Table 3. SkyCalc outputs

| | |
|---|---|
| Target elevation | 32° above horizon (55° peak) |
| Moon elevation | 18° above horizon (48° from target) |
| Sun elevation | 36° below horizon |
| Airmasses | 1.9 |
| Noise spectral radiance | 2 MJy/sr (400 γ/s/m$^2$/μm/as$^2$) |
| Atmospheric transmittance | >80% |

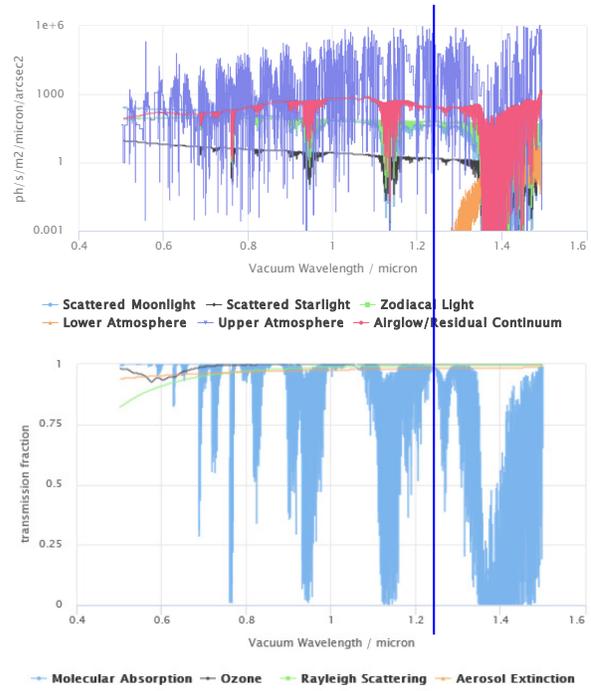

Fig. 3. Top: Radiance components at the telescope site. Bottom: Atmospheric transmittance of the path from the direction of αCen to the telescope site.

#### 3.2.2 Dust

The dust model assumes that αCen A's dust disk is similar to that of the Solar System and contains 1 zodi of dust [10]. To compute the dust radiance as seen from Earth, the central smooth cloud of the DIRBE model [11] has been implemented, re-centered about αCen A, and re-oriented to αCen's orbital plane. The result is shown in Fig. 4.

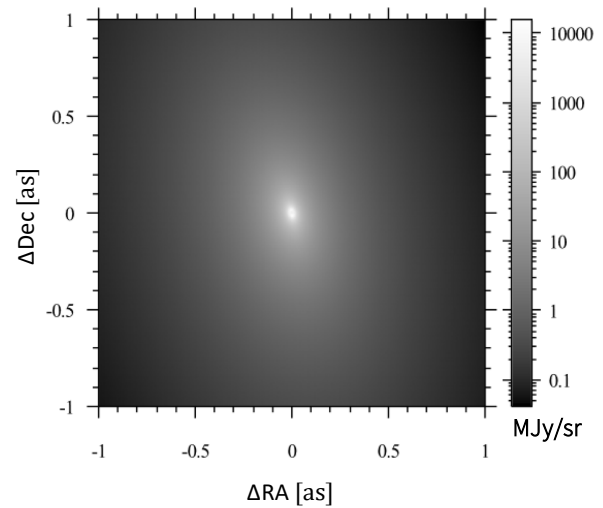

Fig. 4. Noise spectral radiance from αCen A starlight that is reflected and reradiated from dust.

Close to αCen A, the dust can be one or two orders of magnitude brighter than the sky noise. Brightness estimates higher than this, at the center of Fig. 4, are extrapolations that should be regarded with caution, because the DIRBE model was fitted to data at solar elongation angles (angle between the Sun and the line of sight) between 64° to 124° only.

Regardless, these brightness estimates are orders of magnitude dimmer than the source of noise described next. Future modeling will include cases where a coronagraph is used, and in such cases, the dust may become a more important source of noise.

*3.2.3 Direct light*

αCen A has an angular extent of 8 mas ($1.3 \times 10^{-15}$ sr) with a proper motion of 3.7 as/year. In comparison, the 30-meter telescope that tracks the sailcraft's downlink has an 11 mas ($2.2 \times 10^{-15}$ sr) effective field of few. Thus, the star occupies at most 60% of the 30-meter telescope's effective solid angle and does so for less than a day during transit.

αCen A has a noise spectral radiance (brightness) of $3 \times 10^{12}$ MJy/sr at 1.25 μm, as estimated using Planck's law with a black-body temperature of 5,790 K. This is 12 orders of magnitude brighter than the sky and dust.

Unfortunately, once αCen A is no longer in the telescope's field of view, it remains a source of noise. This effect is illustrated by Fig. 5, a digital image taken by the author in Rome. A small amount of the energy going into each pixel is scattered into neighbouring pixels by the atmospheric scattering and the optics of the camera itself. This effect is normally invisible because the scattered irradiance is orders of magnitude below the signal's irradiance. But the sun is many orders of magnitude brighter than its neighbouring objects, so its image spreads to outshine anything close to it.

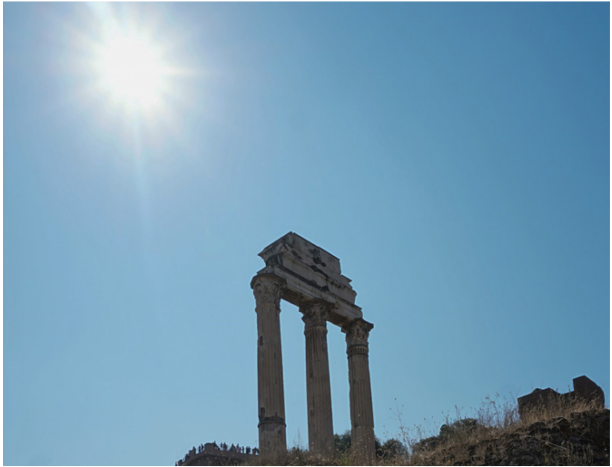

Fig. 5. A simple illustration of how imaging optics scatters bright sources of light into neighbouring pixels.

A Moffat distribution [12] is used to represent the scattering of αCen A's a noise spectral irradiance as a function of angular separation. This distribution can be expressed as

$$\frac{I}{I_0} = \frac{a}{\left[1+\left(\frac{\theta}{\alpha}\right)^2\right]^\beta}. \quad (2)$$

Values of a=0.001, α=0.15 as, and β=1.9 are used to match the seeing-limited point spread function of a 30-meter telescope at 1.25 μm [13]. The resulting point spread function is plotted in Fig. 6. αCen A's spectral irradiance $I_0$ is the product of its spectral radiance and solid angle, and can be expressed as $5 \times 10^7$ γ/s/m²/nm. Expressed as this photon flux, the orders of magnitude fall away as the angular separation increases. Narrowing the spectral filter would also help.

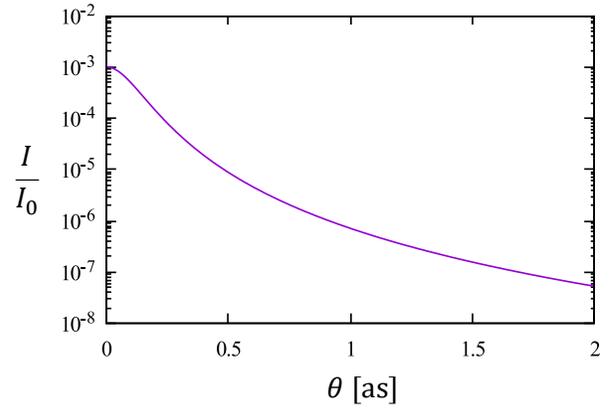

Fig. 6. Seeing-limited point spread function

*3.3 Channel*

The downlink channel is photon-starved, with the 30-meter telescope receiving only 288 signal photons per second. In this regime, pulse position modulation (PPM) is attractive because it is energy efficient, narrowband, and does not require or preclude coherent detection.

The ideal Poisson PPM channel capacity [4] is used as an upper bound for the downlink data rate. It is given by,

$$C = S\eta M \log_2\left(\frac{(1+s)^{q(1+s)} s^{(1-q)s}}{(q+s)^{q+s}}\right) \quad (3)$$

where

$$s = \frac{N}{M \cdot S}, q = \min\left\{\frac{1}{M}, \frac{(1+s)^{1+s}}{es^s} - s\right\} \quad (4)$$

Realized performance only approaches the ideal limit, and high-performance soft-decision schemes are the subject of ongoing research [4]. In lieu of the eventual scheme, a hard-decision Poisson PPM channel model [14] is used to bracket the lower-bound performance of the eventual soft-decision scheme. The hard-decision channel capacity is given by,

$$C = \frac{1}{M}\left[\log_2 M + (1-P_s)\log_2(1-P_s) + P_s \log_2\left(\frac{P_s}{M-1}\right)\right] \quad (5)$$

Probability of symbol error $P_s$ is given by

$$P_s = 1 - \frac{e^{-(n_s+Mn_n)}}{M}$$
$$- \sum_{k=1}^{\infty} \frac{(n_s+n_n)^k e^{-(n_s+n_n)}}{k!}\left[\sum_{m=0}^{k} \frac{n_n^m e^{-n_n}}{m!}\right]^{M-1} \frac{(1+a_k)^{M-1}}{a_k M} \quad (6)$$

where

$$a_k = \frac{\frac{n_n^k}{k!}}{\sum_{m=0}^{k-1} \frac{n_n^m}{m!}} \quad (7)$$

A raw bit error rate of $P_b = P_s M/(2M-2) = 10^{-3}$ is chosen to match the original choice of Lesh et al. [2], who assert that a rudimentary coding scheme reduces this to well below $10^{-6}$. In the present implementation of the hard-decision model, $n_s$ is varied such that the bit error rate matches the desired value. Choosing a value for $n_s$ implies values for dependent variables $n_n$ and $T_s$. Similar to Lesh et al. [2], $M = 1024$ is assumed in all cases here.

## 4. Downlink performance

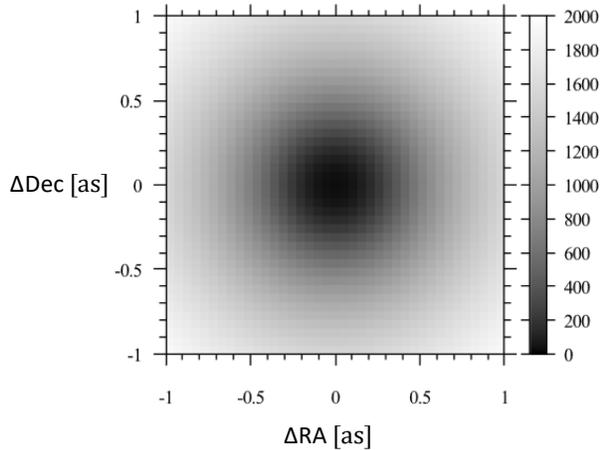

Fig. 7. Ideal channel capacity [bit/s]

In Fig. 7, the ideal channel capacity in Equation (3) is plotted by position in the sky relative to αCen A. The plot shows how direct light scattered by the atmosphere and telescope optics dominates the noise. It reduces the ideal channel capacity in the vicinity of αCen A down to a minimum of 22 bit/s at its center, assuming that αCen is not directly in the receiver's field of view. At 1 as angular separation, ideal channel capacity is 1500 bit/s.

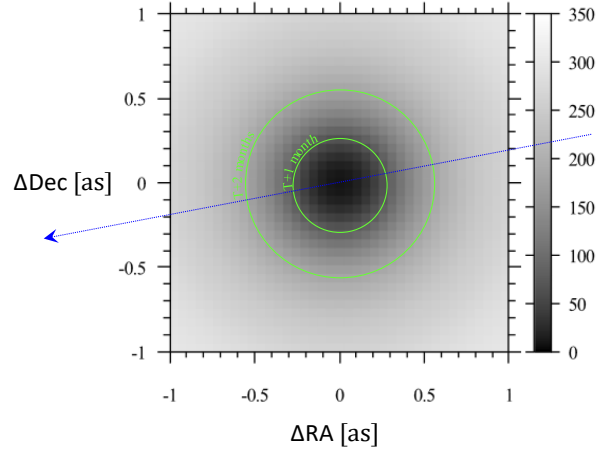

Fig. 8. Hard-decision channel capacity [bit/s]. Sailcraft position due to proper motion of αCen (3.7 as/year) is also shown.

In Fig. 8, the hard-decision channel capacity in Equation (5) is plotted by position in the sky relative to αCen A. This plot also shows how direct light scattered by the atmosphere and telescope optics dominates the noise. It reduces the hard-decision channel capacity in the vicinity of αCen A down to a minimum of 7 bit/s at its center, assuming that αCen is not directly in the receiver's field of view. At 1 as angular separation, the hard-decision channel capacity is 260 bit/s.

## 5. Conclusions

Subject to the assumptions made, each Starshot sailcraft can return 8-50 Gbit/year of raw data from its flyby of αCen A, more than enough to look for signs of life by imaging planets and gathering other scientific data.

If the planned flyby rate of one sailcraft per week is realized, the cumulative pipeline of data will be vast indeed.

Mesh links between sailcraft add reliability and capacity, but they also enable a distributed algorithm operating through the sailcraft before and after their flybys. This distributed algorithm would have a decision-act cycle of a week, as opposed to 9 years for the human decision-act cycle. If the sailcraft have enough cross-range, this provides the basis for an automated exploration of the system in which new planets and moons are first spotted, then their orbits characterized, then observed at close range, then mapped and monitored through successive passes according to human-tended priorities.

Future work needs to retire the leading uncertainties associated with the downlink, which are:
1. Sailcraft aperture efficiency (unknown)
2. Sailcraft available laser power (5 orders of magnitude – 10 mW to 700 W)
3. Receiving telescope filter bandwidth (4 orders of magnitude – 0.1 nm to 10 fm)
4. Receiving telescope size (2 orders of magnitude – 1 meter to 100 meters)
5. PPM downlink coding scheme (1 order of magnitude – 1 γ/bit to 10 γ/bit)
6. Receiving telescope point spread function / adaptive optics / coronagraph (unknown, assumed ~ 1 order of magnitude)
7. Receiving telescope filter insertion loss (<1 order of magnitude)
8. Choice of wavelength (assumed <1 order of magnitude effect)

Also, the next downlink model should:
- Include the effect of a coronagraph
- Include astrometric positions taking into account orbits of the Earth, Sun, and αCen A & B
- Consider how multiplexing is implemented for a pipeline of sailcraft
- Investigate smaller, geographically diverse, high dynamic range telescopes to minimize the probability of missed downlink data and system cost

**Acknowledgements**

This work was supported by the Breakthrough Prize Foundation. I thank Professors Avi Loeb, Phil Lubin, David Messerschmitt, and Olivier Guyon for their most helpful feedback.